\documentclass[11pt,a4paper]{article}

\usepackage[utf8]{inputenc}
\usepackage[T1]{fontenc}
\usepackage{lmodern}
\usepackage[a4paper,margin=2.4cm]{geometry}

\usepackage{graphicx}
\usepackage{amsmath}
\usepackage{booktabs}
\usepackage{authblk}
\usepackage{fancyhdr}
\usepackage{caption}
\usepackage{subcaption}
\usepackage{setspace}
\usepackage{microtype}
\usepackage[numbers,sort&compress]{natbib}

\newcommand{\threesubsection}[1]{\subsection*{#1}}

\providecommand{\citen}[1]{\cite{#1}}

\title{\textbf{Bias-Engineered Synthetic Antiferromagnets Hosting sub-20 nm Zero-Field Skyrmions at Room Temperature}}

\author[1,*]{Emily Darwin}
\author[2]{Riccardo Tomasello}
\author[1,4]{Reshma Peremadathil Pradeep}
\author[2]{Mario Carpentieri}
\author[3]{Giovanni Finocchio}
\author[1,4,*]{Hans J. Hug}

\affil[1]{Empa, Swiss Federal Laboratories for Materials Science and Technology, Ueberlandstrasse 129, 8600 Dübendorf, Switzerland}
\affil[2]{Department of Electrical and Information Engineering, Politecnico di Bari, I-70125 Bari, Italy}
\affil[3]{Department of Mathematical and Computer Sciences, Physical Sciences and Earth Sciences, University of Messina, I-98166 Messina, Italy}
\affil[4]{Department of Physics, University of Basel, Klingelbergstrasse 82, 4056 Basel, Switzerland}
\affil[*]{Corresponding authors: emily.darwin@empa.ch; hans-josef.hug@empa.ch}

\date{}

\begin{document}

\maketitle

\begin{abstract}
Synthetic antiferromagnetic skyrmions (SAFsk) are nanoscale, topologically protected spin textures with strong potential for spintronic technologies because of their high stability and the absence of the skyrmion Hall effect. However, robust zero-field stabilization remains a central challenge. Here, a synthetic antiferromagnetic (SAF) bias system is introduced as a novel strategy to stabilize both ferromagnetic skyrmions (FMsk) and SAFsk at zero field. Ferromagnetic (FM) and SAF multilayers are designed, fabricated and integrated with the SAF bias system to enable controlled skyrmion stabilization and polarity setting via multilayer design and a preparatory field cycle. Combining quantitative and high-sensitivity magnetic force microscopy (MFM) with micromagnetic modeling, reliable zero-field skyrmion formation is demonstrated and sub-20\,nm SAFsk are directly observed, the smallest SAFsk reported to date. Moreover, the SAF bias system concept introduced here offers a robust and scalable route to bias future skyrmion multilayers, as its compensated nature suppresses domain formation and preserves a uniform exchange field.
\end{abstract}
\flushbottom

\newpage
\thispagestyle{empty}

\section{Introduction}
Magnetic skyrmions are nanoscale, particle-like configurations of spins, stabilized by a trade-off among different magnetic energies with a nontrivial topology.\cite{finocchio2016magnetic,rodrigues2025skyrmions,back20202020} They can be observed in sputtered ferromagnetic (FM) multilayers (ML) with the presence of the Dzyaloshinskii–Moriya interaction (DMI).

Owing to their topological stability and well-defined dynamics, skyrmions are considered promising candidates for information carriers in storage and racetrack devices and unconventional applications.\cite{dohi2022thin,zazvorka2019thermal,song2020skyrmion,beneke2024gesture,li2021magnetic,koraltan20262026}
For technological applications, achieving device compactness and scalability requires skyrmions with ultra-low diameters. In racetrack devices, lateral confinement promotes alignment into one-dimensional chains;\cite{du2015edge} while skyrmions must be sufficiently spaced to allow reliable motion and detection, reducing their size remains crucial, as it sets the minimum achievable bit spacing and thus the maximum storage density. Current skyrmion sizes in FM ML without lateral confinement reach down to around 50 nm.\cite{soumyanarayanan2017tunable,raju2019evolution,ho2019geometrically} 
However, further size reduction remains difficult because ferromagnetic skyrmions (FMsk) are subject to strong dipolar interactions that set a lower limit to their diameter.\cite{buttner2018theory} In addition, current-driven transport is hindered by the skyrmion Hall effect (SkHE),\cite{jiang2017direct,litzius2017skyrmion} which causes transverse deflection and edge annihilation. To address these challenges, synthetic antiferromagnetic (SAF) ML have been developed as a material platform for stabilizing and manipulating skyrmions.\cite{legrand2020room,dohi2019formation,juge2022skyrmions,pham2024fast,darwin2024antiferromagnetic,chen2020realization}

A SAF consists of FM layers coupled through interlayer exchange coupling (IEC) across a non-magnetic spacer. Certain spacer materials, such as ruthenium, exhibit an oscillatory IEC behavior.\cite{parkin1991systematic} By varying the spacer layer thickness, the coupling can transition between FM and antiferromagnetic (AFM). Systems that exhibit AFM-IEC are referred to as SAFs, and have already served as key building blocks in device architectures such as spin valves and magnetic tunnel junctions.\cite{parkin2015memory,liu2025topological} For skyrmion-based devices, SAFs can be advantageous as the AFM-IEC compensates dipolar fields allowing for a smaller skyrmion size and suppresses the SkHE,\cite{zhang2016magnetic,tomasello2017performance} enabling fast, straight skyrmion motion.\cite{barker2023domain,yang2015domain,alejos2018current,dohi2019formation,pham2024fast}

In the absence of an external magnetic field, SAFs are typically in a uniform AFM configuration, for example, with the magnetization of two antiferromagnetically coupled FM layers aligned antiparallel. When the DMI is present, however, narrow antiparallel maze domain patterns can emerge.\cite{legrand2020room} These non-uniform states increase the overall Zeeman energy due to local opposing magnetizations in the two layers. The micromagnetic state of a SAF thus depends on a delicate balance among the DMI, IEC, and Zeeman energies.

Individual SAF skyrmions (SAFsk) can emerge as localized reversal states within an otherwise homogeneous background,\cite{sim2025zero} although their spontaneous nucleation is rare. While previous studies have shown that SAFsk can be generated at current injection points,\cite{juge2022skyrmions} or induced through magnetic field sequences,\cite{pham2024fast,dohi2019formation} their size and stability in antiferromagnetically coupled systems remain largely unexamined. Establishing control over the formation and stability of small, well-defined SAFsk under zero-field conditions is therefore crucial for their incorporation into scalable spintronic devices. One approach to realizing small, zero-field SAFsk involves the use of a FM bias layer.\cite{legrand2020room} Typically, an applied magnetic field is required for stabilizing skyrmions, however, zero-field skyrmions can be achieved by incorporating a FM bias layer beneath a skyrmion hosting multilayer. This layer exerts an internal exchange field on the adjacent lowermost magnetic layer, effectively replacing the need for an external field.\cite{legrand2020room}

While exchange coupling and bias layer engineering have been extensively investigated in FMsk systems,\cite{ajejas2023densely,chen2024tailoring,brandao2022tuning} their role in SAF multilayer systems remains far less explored, primarily because direct observation of SAFsk is inherently challenging. Conventional techniques that probe the net magnetic moment averaged through the film thickness are largely insensitive to the compensated nature of SAFs, since the opposing magnetization of two magnetic layers effectively cancel out the measurable signal.

To directly observe SAFsk, the finite stray field just above the film surface can be exploited, as the upper magnetic layer is closer to the field-probe and only partially compensated by the lower layer. However, the overall stray field amplitude remains extremely small, as the contributions from both layers nearly cancel at the lift heights typically employed in scanning probe measurements. The small size of SAFsk further reduces the detectable signal, necessitating imaging techniques with exceptional magnetic sensitivity. Consequently, their experimental detection relies on advanced stray field-based measurement methods, such as quantitative magnetic force microscopy (qMFM), capable of resolving the minute contrast associated with these compensated textures. Despite some studies imaging SAFsk via MFM, control of skyrmion polarity is yet to be achieved, and most experimental observations to date report relatively large skyrmion diameters ($>$50 nm).\cite{legrand2020room,sim2025zero,juge2022skyrmions}

In this work, we present a refined SAF bias system concept and demonstrate its potential to enhance system robustness and enable controlled zero-field skyrmion stabilization, offering a viable design pathway for future SAFsk technologies. 
We fabricate two different systems: one FM ML and one SAF ML, both utilizing the same SAF bias system to achieve zero-field skyrmion stabilization with a polarity tunable through system design and a preparatory magnetic field cycle. Using high-sensitivity qMFM\cite{feng2022quantitative} and micromagnetic simulations, we confirm that the SAFsk achieved here are substantially smaller than their FM counterparts, reaching sub-20 nm diameters, the smallest SAFsk observed to date. Our results establish a robust route toward controlled, zero-field SAFsk formation and highlight the efficacy of bias layer engineering for advancing skyrmion-based device concepts with ultra-low energy consumption.

%========================
% 3a. bias system concept and motivation
%========================
\section{Design of the Systems}
\subsection{Strategy for the bias system design}
A bias layer is designed to provide a homogeneous exchange field acting on the bottommost magnetic layer of a skyrmion-supporting multilayer. However, when the energy required to stabilize skyrmions in the skyrmion ML exceeds the energy cost of forming domain walls in the bias layer, the bias layer itself becomes unstable against domain formation. In this case, the skyrmion ML can remain in its zero-field domain state and imprint this configuration into the bias layer.

FM bias layers are typically Pt/Co multilayers with low Co thickness and few repetitions to maintain a single-domain state. Increasing magnetic anisotropy or the number of Pt/Co repetitions can partially mitigate the instability, but the latter cannot be increased arbitrarily, beyond a certain repetition number, the rise of magnetostatic energy promotes domain formation compromising uniform biasing (see Supporting Information Section 2 for a more detailed analysis of the stability requirements for the biasing layer).

To overcome these limitations, we employ a SAF as the bias layer. In principle, the SAF could also form domains, with any domain in one coupled layer necessarily mirrored by an oppositely magnetized domain in the other layer. Such configurations are energetically unfavorable because the stray field of a domain in one layer is antiparallel to the magnetization of the corresponding domain in the other layer, which produces a large magnetostatic (Zeeman) penalty. This high energy cost effectively suppresses domain formation in the SAF and keeps it uniformly magnetized. 

In contrast to exchange-bias based stabilization schemes, where pinning from interfacial spin disorder can severely hinder motion,\cite{schmid2010exchange,benassi2014role,rana2020room,takano1997interfacial,stiles1999model} the RKKY-mediated effective field in the present SAF design originates from a homogeneously magnetized reference layer and is therefore expected to provide a smoother energy landscape. This suggests that SAF-based biasing may overcome the commonly observed trade off between zero-field stability and skyrmion mobility.

%========================
% 3b. SAF bias system implementation and magnetometry
%========================
\subsection{SAF bias system design}
\begin{figure*}[thb!]
    \centering
    \includegraphics[width=1\textwidth]{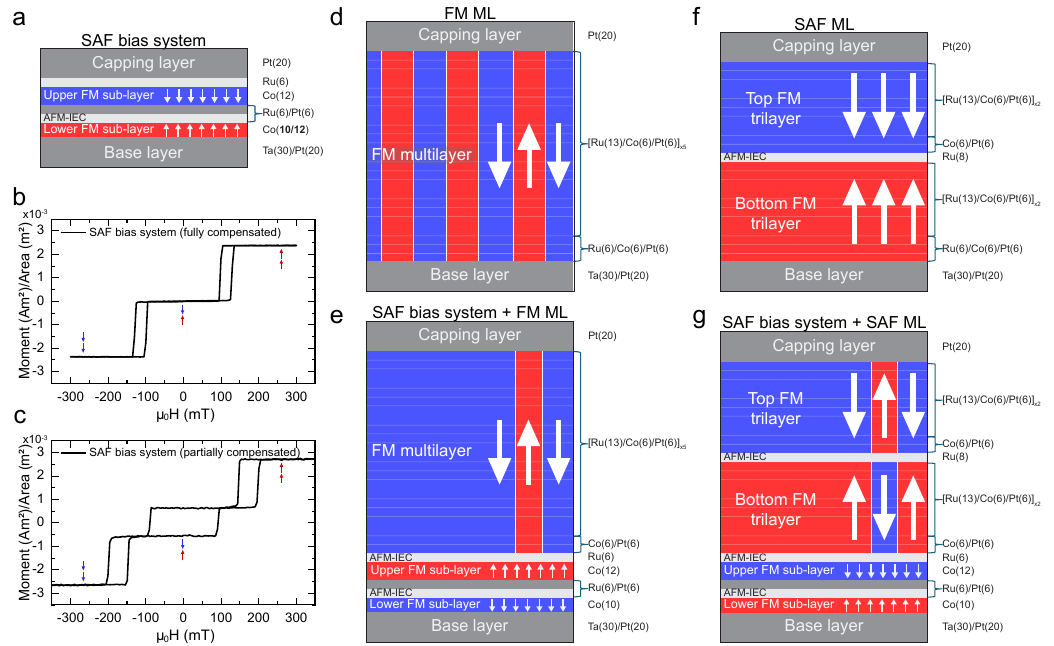}
    \caption{a Schematic of the SAF bias system, layers providing the AFM-IEC are indicated. b Hysteresis loop of the fully compensated SAF bias system. c Hysteresis loop of the partially compensated SAF bias system. Red/blue arrows indicate magnetization switching events. d Schematic of the FM ML. e SAF bias system coupled with the FM ML. f Schematic of the SAF ML. g SAF bias system coupled with the SAF ML. White arrows indicate the expected local magnetic moment orientations after saturation in a south field and returned to zero field. All thicknesses are in~\AA.}
    \label{fig:1}
\end{figure*}

All material systems were grown by magnetron sputtering (see Experimental Section). Each system uses a 50~\AA\ Ta and 20~\AA\ Pt seed layer, and a 20~\AA\ Pt cap for oxidation protection. The SAF bias system serves as a general exchange-field source for both FMsk and SAFsk. It consists of a 12~\AA\ upper Co layer and a lower Co layer of either 12~\AA\ or 10~\AA\, separated by a Ru (6~\AA)/Pt (6~\AA) interlayer that mediates AFM RKKY IEC. In the bias system, Pt lies beneath Co and Ru above, a configuration that promotes high-quality growth and maintains strong perpendicular magnetic anisotropy (PMA) even for relatively thick Co (Supporting Information Section 1). Two variants were constructed: a \textit{fully compensated} system and a \textit{partially compensated} system, achieved by setting the lower layer Co thickness to 12~\AA\ or 10~\AA, respectively. The partially compensated design enables control of skyrmion polarity in the overlying ML.

Figure~\ref{fig:1}a shows the SAF bias system. If the partially compensated bias system is saturated in a south field and returned to zero field, the upper Co layer is expected to remain south while the lower Co flips north via AFM-IEC across the Ru, forming the SAF; in the fully compensated case, either sub-layer may switch. Out-of-plane hysteresis loops taken via vibrating sample magnetometry (VSM) for the compensated and partially compensated systems (Figures.~\ref{fig:1}b and c) show sharp transitions between the zero-field SAF state and the high-field FM state.

%========================
% 3c. Multilayer architecture and coupling design
%========================
\subsection{Multilayer and coupling design}
In the skyrmion-supporting ML, Ru is placed beneath Co, and Pt above, which lowers the PMA (Supporting Information Section 1). This enables skyrmion formation at lower fields, ensuring that the energy required for their stabilization remains below that needed to induce domains in the SAF bias system.

\textit{FM ML}: Figure~\ref{fig:1}d shows the FM ML, which was first examined to assess whether the SAF bias system could induce the formation of FMsk that are easier to observe. The FM ML consists of one repetition of Ru(6~\AA)/Co(6~\AA)/Pt(6~\AA) followed by five repetitions of Ru(13~\AA)/Co(6~\AA)/Pt(6~\AA). The Ru thickness of the first layer was kept thinner (6 instead of 13~\AA) to remain consistent with the SAF bias system + FM ML system in Figure~\ref{fig:1}e.  The FM ML system was designed to promote skyrmion formation by establishing a clockwise DMI through the asymmetric placement of Ru and Pt layers around the Co layers, and by optimizing the layer thicknesses to achieve favorable FM-IEC and PMA. If the system is saturated in a south (or north) field and then returned to zero field, it is expected to be in a maze domain state. 

\textit{FM ML on SAF bias system}: Figure~\ref{fig:1}e shows a schematic of the FM ML coupled antiferromagnetically to the SAF bias system beneath it via 6~\AA\ of Ru, designed to stabilize FMsk at zero-field. If the system is saturated in a south field, all magnetic layers would initially align south. As the applied field is reduced to zero, and  drops below the threshold required to maintain the RKKY coupling, the system relaxes to its lowest-energy configuration. This involves a reversal of the upper Co layer of the bias SAF to the north-oriented state, thereby satisfying the AFM-IEC across all layers. Hence, the FM ML is expected to remain nearly uniformly magnetized south at zero field, provided that the RKKY exchange field from the upper north-oriented Co layer of the bias SAF is sufficiently strong, with the possible presence of north-core skyrmions. Conversely, if the system would be saturated in a north field, the opposite scenario occurs, resulting in south-core skyrmions. This thus provides an effective way to control the FMsk polarity.

\textit{SAF ML}:
To achieve and experimentally detect SAFsk, we designed a fully compensated SAF ML that is both capable of hosting skyrmions and optimized for their magnetic stray field detection. The main experimental challenge arises from the compensated nature of SAFs: the opposing magnetization in the two sub-layers yields a vanishing net magnetic moment when integrated through the film thickness. Consequently, imaging techniques that are sensitive to the total magnetic moment, such as Lorentz transmission electron microscopy (LTEM) or X-ray magnetic circular dichroism (XMCD)–based scanning transmission X-ray microscopy (STXM), cannot easily detect compensated textures like SAFsk.\cite{juge2022skyrmions,zhao2023identifying,christensen20242024} Even emerging three-dimensional XMCD tomography methods, while promising, currently lack sufficient spatial resolution along the film normal to resolve the magnetic structure of ultra-thin ($\approx 2\,$ nm) SAF layers.\cite{hierro2020revealing,donnelly2020imaging}

Instead, methods such as MFM, which detect magnetic textures indirectly through the stray fields emanating from the sample surface, can be utilized. The amplitude of these fields decays approximately as $e^{-k z}$, where $z$ is the probe–sample distance and $k = 2\pi/\lambda$ is the spatial wave vector corresponding to the magnetization wavelength $\lambda$.\cite{hug1998quantitative} Therefore, in a SAF, the stray field from the lower magnetic layer is attenuated more strongly than that from the upper layer, leading to an incomplete cancellation of their opposing contributions at the probe height. This slight imbalance allows SAFsk to produce a small but finite MFM contrast. For SAF systems consisting of only 2 nm-thin magnetic layers separated by a $\sim$1 nm non-magnetic spacer which provides the AFM RKKY coupling, this differential signal becomes exceedingly weak—likely below the detection limit of conventional, lower-sensitivity MFM instruments operating under ambient conditions. Even with high-sensitivity qMFM measurements using high-quality factor ($Q$) cantilevers under vacuum, the remaining $\Delta f$ contrast is expected to be extremely small and may be difficult to distinguish from background frequency-shift variations that arise from surface roughness via van der Waals interactions or from local variations in magnetic moment.\cite{feng2022magnetic}

To overcome these limitations and make SAFsk experimentally observable, we designed the SAF ML as two ferromagnetically coupled trilayers, [Ru/Co/Pt]$_{\times 3}$, separated by an 8~\AA\ Ru spacer mediating AFM-IEC. This architecture increases the effective magnetic thickness of the top and bottom trilayers and thus enhances the relative contribution of the top trilayer to the stray field at the probe height, thereby improving detectability while preserving the compensated nature of the SAF system.

Figure~\ref{fig:1}f illustrates the optimized SAF ML used; it shares the same [Ru/Co/Pt] design as the FM ML but utilizes a thinner central Ru spacer of 8~\AA\ instead of 13~\AA. This 
configuration not only enables the formation of SAFsk but also enhances their detectability in stray field-based measurements, as discussed above. Being fully compensated, the two SAF-coupled trilayers have no preferred magnetization orientation at zero field after saturation in a south (or north) field, other than being antiparallel to each other.

\begin{figure*}[th!]
\centering
\includegraphics[width=1.0\textwidth]{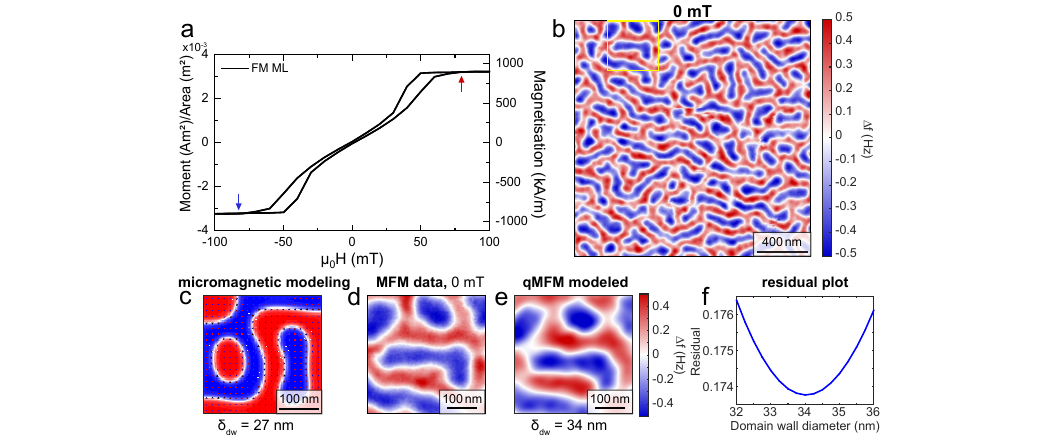}
\caption{a VSM hysteresis loop of the FM ML. b Background-subtracted MFM data of the FM ML at 0\,mT. c Micromagnetic simulation of the FM ML revealing a domain wall width of $\delta_{\rm dw} = 27\,$nm. d Magnified area of the MFM data in b, highlighted by the yellow box. e Simulated  MFM $\Delta f$-contrast with a domain wall width of $\delta_{\rm dw} = 34\,$nm, optimized to match the experimental data shown in d. f Dependence of the residual rms-difference between the simulated and measured $\Delta f$-contrast on the domain wall width revealing a pronounced minimum at 34\,nm.}
\label{fig:2}
\end{figure*}

\textit{SAF ML on SAF bias system}:
Figure~\ref{fig:1}g illustrates the SAF ML coupled antiferromagnetically to the SAF bias system beneath it via 6~\AA\ of Ru, designed to stabilize a zero-field SAFsk state in the SAF ML. After saturation in a south field, all magnetic layers initially align south. 
Upon removing the applied field, the system relaxes toward its lowest-energy configuration through a sequence of AFM-IEC-driven switching processes.
By using an only partially compensated SAF bias system, the thicker, upper Co layer in the SAF bias system is expected to remain south-oriented, while the thinner lower Co layer switches north to satisfy the internal AFM coupling.
This, in turn, imposes an AFM exchange field on the overlying SAF ML, aligning its lower three Co layers predominantly north, potentially hosting south-core skyrmions.
To satisfy the AFM-IEC in the SAF ML, the top three Co layers would be oriented nearly uniformly south, with the possible presence of north-core skyrmions. Conversely, if the system were saturated in a north field, the opposite scenario occurs, providing a possible way to control the orientation of the SAF layers, and therefore skyrmions, by saturating the system in a north or south field.

%-----------------------------------------------------------
%-----------------------------------------------------------
%-----------------------------------------------------------

\section{Results}
In the following subsections, we use magnetometry to determine the overall magnetic response and switching behavior of each system, defining the field regimes relevant for skyrmion formation. High-sensitivity MFM and qMFM modeling are performed to obtain and analyze the characteristic micromagnetic textures. Micromagnetic simulations are used to determine the typical spin configurations at zero field and to provide input for the quantitative modeling of the MFM measurements. We first analyze the FM ML to establish and validate the qMFM approach, followed by investigations of the SAF ML, and the skyrmion spin textures in the SAF bias system + FM ML and the SAF bias system + SAF ML systems.\\

To suppress spurious contrast arising from surface roughness or local variations in magnetic moment, all MFM images in a non-uniform state were background-subtracted by subtracting data acquired either in magnetic saturation or at fields slightly above the skyrmion collapse field (see Supporting Information Section 3).\cite{bacani2019measure,feng2022magnetic} The resulting background-subtracted images therefore reveal only the magnetic contrast associated with domain structures and skyrmion textures. 
For all systems, micromagnetic simulations were performed to produce the zero-field states of the multilayer systems and used measured material parameters: saturation magnetization ($M_{\rm s}$), uniaxial anisotropy ($K_{\rm u}$) and RKKY coupling strength ($J_{\rm RKKY}$). Plus, a reasonable exchange stiffness ($A$) and DMI constant ($D$). The values are shown in the Experimental Section and in Table \ref{table:mag_parameters_micro}.\\

To achieve a quantitative comparison between simulated and experimental MFM data, the sensitivity of the MFM tip to magnetic stray fields of different spatial wavelengths was calibrated following the established procedure in Reference~\citen{feng2022quantitative}, using a  Si/Pt(100)/[Co(6)/Pt(10)]$_{\times 6}$/Pt(30) multilayer reference sample. The resulting tip transfer function enables a quantitative calculation of simulated MFM frequency-shift maps ($\Delta f$) from the stray fields generated by candidate magnetization patterns. The magnetization patterns, initially guided by the micromagnetic simulations, were iteratively refined by adjusting parameters such as skyrmion radius and domain wall width until quantitative agreement between simulated and experimental frequency-shift data was reached. Details of the modeling formalism and fitting procedure are given in the Experimental Section.
The domain wall width can be extracted with particularly high precision. Clockwise Néel walls, arising from the Ru/Co/Pt design of our multilayer systems that generate negative DMI, lead to magnetic volume charges that partly attenuate the stray field signal from surface charges. The existence of these volume magnetic charges make the resulting MFM contrast highly sensitive to nanometer-scale variations in wall width. This methodology, combining high-sensitivity background-corrected MFM, calibrated tip response, and quantitative magnetization modeling, enables reconstruction of nanoscale skyrmion textures, including their true radii and domain wall widths, with sub-nanometer accuracy.

\subsection{Ferromagnetic Multilayer}
To establish and validate both the micromagnetic modeling and the quantitative simulation of MFM data, we first analyze the FM ML by VSM and high-sensitivity MFM. The VSM hysteresis loop (Figure~\ref{fig:2}a) exhibits the typical linear and nearly hysteresis-free magnetization behavior up to about 30~mT, followed by slightly hysteretic skyrmion pockets between 30-70~mT, and saturation above 70~mT. The background-subtracted MFM image of the FM ML at zero field (Figure~\ref{fig:2}b) reveals the characteristic maze domain pattern expected for DMI-stabilized systems, with an average domain periodicity of $(133\pm 32)$~nm as determined from Fourier analysis. The reduced domain size is consistent with the presence of a finite DMI, which lowers the domain wall energy.\\
The corresponding micromagnetic simulation (Figure~\ref{fig:2}c) reproduces the experimentally observed domain morphology, yielding clockwise Néel walls consistent with the negative DMI expected for the Ru/Co/Pt interface design. The modeled domain wall width is approximately $\delta_{\text{dw}} = 27~\text{nm}$. 
Quantitative MFM modeling was performed by simulating the MFM signal generated from a magnetization structure composed of alternating north/south domains with $M_{\rm s}=920\,$ kA/m, separated by clockwise Néel domain walls of finite width. The magnetization pattern was derived by discriminating the measured frequency-shift map and introducing domain walls with widths between 32 and 36~nm to reproduce the observed contrast. A best-fit comparison between the experimental $\Delta f$ (Figure~\ref{fig:2}d) and the simulated data  (Figure~\ref{fig:2}e) yields an optimal domain wall width of $\delta_{\text{dw}} = (34.0\pm 0.5)~\text{nm}$, in close agreement with the micromagnetic result.
The robustness of this fitted value was verified by calculating the residual rms-difference between simulated and experimental $\Delta f$ maps for varying wall widths (Figure~\ref{fig:2}f). The residual curve exhibits a distinct minimum near $\delta_{\text{dw}} = 34.0~\text{nm}$, confirming the high precision of the qMFM analysis. The excellent agreement among micromagnetic modeling, experimental MFM data, and quantitative MFM simulation thus validates the accuracy and reliability of the combined analysis approach.

\begin{figure*}[t]
\centering
\includegraphics[width=1.0\textwidth]{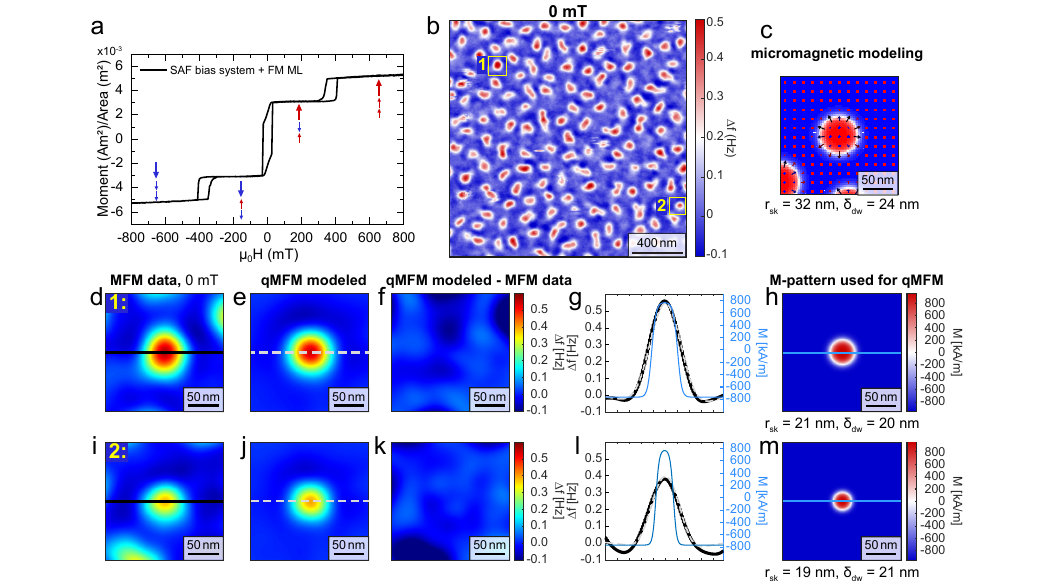}
\caption{a VSM hysteresis loop of the SAF bias system + FM ML. b Background-subtracted MFM data of the SAF bias system + FM ML at 0 mT. c Micromagnetic simulation of the SAF bias system + FM ML showing a FMsk stabilized through the SAF bias system at zero field. Analysis of skyrmion 1 (d–h) and skyrmion 2 (i–m): d,i Background-subtracted MFM data cut from the overview image displayed in b at the locations highlighted by the yellow boxes and marked by the labels 1 and 2. e,j qMFM simulation of the skyrmions with optimized radii and domain wall diameters. f,k Residual maps from subtracting the experimental from the modeled data. g,l Line profiles of the experimental (solid black lines) and the modeled (dashed gray lines) MFM skyrmion frequency-shift signals and skyrmion magnetization profiles (blue lines). h,m Optimized magnetization patterns used for qMFM analysis of the skyrmion.}
\label{fig:3}
\end{figure*}

\subsection{Ferromagnetic Multilayer on SAF bias system}
The partially compensated SAF bias system was incorporated beneath the FM ML to provide an RKKY exchange field capable of stabilizing skyrmions in the absence of an external magnetic field. The corresponding hysteresis loop, shown in Figure~\ref{fig:3}a, exhibits two distinct switching processes. The inner, narrow loop centered around zero field corresponds to a magnetization reversal at approximately $\pm 30~\text{mT}$, while a second, outer switching occurs at approximately $\pm 380~\text{mT}$. The latter transition is associated with an abrupt change in magnetic moment per unit area of roughly $2\times10^{-4}~\text{emu/cm}^2$, which matches the switching of the SAF bias system between its antiparallel and parallel magnetic configurations (see Figure~\ref{fig:1}b and c), albeit at significantly higher fields than the isolated SAF bias system, which switches at around $\pm120~\text{mT}$.

This increase in switching field can be understood from the magnetic configuration illustrated in Figure~\ref{fig:1}e, which represents the system state at zero field after saturation in a south-directed field. In this configuration, only the upper Co layer of the bias SAF points north, while both the FM ML and the lower Co layer of the bias SAF remain oriented south. Upon reducing the applied field after south saturation, the AFM RKKY coupling across both the Ru(6~\AA) spacer above and the Ru(6~\AA)/Pt(6~\AA) spacer layers below cooperatively act to flip the upper Co layer of the bias SAF northward, counter to the still-applied external field. This cooperative coupling effect explains the elevated switching field observed for the SAF bias system when coupled to the FM ML.

The change in moment near $\pm30~\text{mT}$ closely corresponds to that observed for the FM ML alone (Figure~\ref{fig:2}a) and is therefore attributed to the reversal of the FM ML, accompanied by a simultaneous reversal of the magnetization in both sub-layers of the SAF bias system.
From the hysteresis loop shown in Figure~\ref{fig:3}a, we deduce that after saturation in a south-directed magnetic field, the majority of magnetic moments in the FM ML remain oriented south at zero applied field. The slight reduction in the net south-directed moment observed at zero field is consistent with the nucleation of skyrmions possessing north-oriented cores.
MFM imaging, shown in Figure~\ref{fig:3}b, directly confirms the presence of these skyrmions at zero applied field. The observed positive frequency-shift corresponds to a repulsive magnetic interaction between the north-core skyrmions and the MFM tip, which has a south-oriented magnetization (i.e., a north pole facing the sample surface). 
Together, the hysteresis analysis and the MFM contrast demonstrate that the SAF bias system produces a robust and spatially uniform internal exchange field that stabilizes a well-ordered skyrmion lattice in the FM ML at zero external field, with the potential to select the skyrmion core polarity depending on whether the system is first saturated in a south or north field.

\begin{figure*}[ht]
\centering
\includegraphics[width=1.0\textwidth]{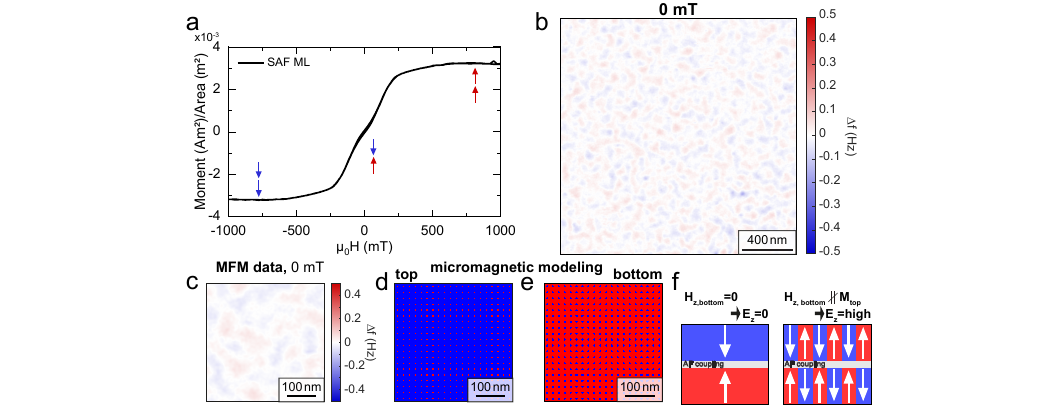}
\caption{a VSM hysteresis loop of the SAF ML. b MFM data of the SAF ML at 0\,mT revealing only background contrast but no skyrmions. c MFM data of the SAF ML at 0\,mT on the same scale as the modeled data. d Micromagnetic simulation of the top trilayer of the SAF ML. e Micromagnetic simulation of the bottom trilayer of the SAF ML. f Schematic of magnetization configurations with low and high Zeeman energy (left and right panels respectively).}
\label{fig:4}
\end{figure*}

Figure~\ref{fig:3}c presents a micromagnetic simulation of the system in a relaxed state without an applied field. It shows a stabilized FMsk, with a radius of $r_{\text{sk}} = 32~\text{nm}$ and a domain wall width of $\delta_{\text{dw}} = 24~\text{nm}$. Figures~\ref{fig:3}d and i show a single FMsk in the SAF bias system + FM ML system, cut from Figure~\ref{fig:3}b (highlighted by the yellow squares labeled 1 and  2). 

To quantitatively model the measured MFM frequency-shift contrast of individual skyrmions (Figure~\ref{fig:3}d–m), we use a parametrized circularly symmetric magnetization profile defined by a skyrmion radius $r$ and domain wall width $\delta_{\mathrm{dw}}$, guided by the relaxed configuration obtained from micromagnetic simulations (Figure~\ref{fig:3}c). Iterative adjustment of $r$ and $\delta_{\mathrm{dw}}$ yields the optimized magnetization patterns in Figure~\ref{fig:3}h and m, which best reproduce the experimental $\Delta f$ contrast displayed in Figure~\ref{fig:3}d and i. The corresponding fitted parameters are $r_{\rm sk}=21$~nm, $\delta_{\rm dw}=20$~nm for the skyrmion in Figure~\ref{fig:3}h, and $r_{\rm sk}=19$~nm, $\delta_{\rm dw}=21$~nm for the skyrmion in Figure~\ref{fig:3}m. Figures~\ref{fig:3}e and j show the associated modeled MFM contrast for the skyrmions in Figure~\ref{fig:3}d and i, respectively. The difference images in panels f and k, together with the cross sections in Figure~\ref{fig:3}g and l, demonstrate the excellent agreement between experiment and simulation, confirming that the reconstructed magnetization textures accurately capture the true skyrmion structure.

It is important to note that the skyrmion diameter inferred directly from the frequency-shift images is significantly larger than the actual diameter of the underlying magnetization texture. This discrepancy arises because the stray field emanating from the skyrmion broadens with increasing tip–sample distance, and because the finite width of the magnetic charge distribution of the MFM tip further spreads the measured contrast. These effects become particularly pronounced when the intrinsic skyrmion diameter approaches the lateral extent of the tip’s magnetic charge distribution. The observed variation in skyrmion size is attributed primarily to local variations in magnetic parameters arising from interfacial roughness in the multilayer stack (see Supporting Information Sections 1 and 2), rather than to spatial inhomogeneities of the SAF-induced effective field, which is expected to be relatively uniform.

\subsection{SAF Multilayer}
SAFs with interfacial DMI can host a remarkably diverse range of magnetic ground states. In the simplest case, and in the absence of strong competing interactions, a SAF adopts a homogeneous antiparallel configuration, as commonly exploited in reference layers of magnetic memory structures.\cite{duine2018synthetic} When the DMI becomes sufficiently large, however, the ground state may instead consist of narrow antiparallel maze domains, as reported for Pt/Co/Ru-based SAFs,\cite{legrand2020room} or bubble domains in Pt/Co/CoFeB/Ir ML.\cite{dohi2019formation} Similar skyrmion bubbles with diameters spanning from the micron scale down to below $100~\mathrm{nm}$ have also been observed in Pt/Co/Ru-based SAFs under suitable combinations of IEC, DMI, and field cycles, including the application of in-plane fields.\cite{pham2024fast} This wide variety of states underscores that the micromagnetic configuration of a SAF is highly sensitive to the balance among IEC, DMI, and any internal or external magnetic field. 

Figure~\ref{fig:4}a shows the VSM hysteresis loop of the SAF ML investigated here, which exhibits the expected response of a compensated SAF. In particular, the magnetic moment at zero field vanishes, confirming the AFM-IEC between the top three and bottom three Co layers and demonstrating that the SAF ML system is fully compensated. 
The corresponding zero-field MFM data in Figure~\ref{fig:4}b reveal that the system resides in a uniform SAF state. Only small variations in the magnetic background are visible. 
A magnified area of the MFM data in Figure~\ref{fig:4}c is compared to the micromagnetic modeling results in Figures~\ref{fig:4}d and e. Figure~\ref{fig:4}d displays the top layer of the SAF ML, where all three top layers are in a south magnetic state, whereas Figure~\ref{fig:4}e shows the bottom layer of the SAF ML, where all three bottom layers are in a north magnetic state. 
We attribute the absence of domains or skyrmions in this system to the substantial Zeeman energy ($E_{\mathrm{z}}$) cost associated with forming antiparallel domain patterns in the two antiferromagnetically coupled FM trilayers, which energetically disfavors non-uniform magnetic configurations in this SAF ML. In the uniform state, the stray field is effectively canceled, the positive magnetic charges at the top surface are compensated by the negative charges at the bottom surface. However, if both layers were to form domains, their magnetization would be locally antiparallel. In this case, the domain in the bottom layer would generate a stray field ($H_{\mathrm{z,bottom}}$) inside the top layer that points opposite to the local magnetization ($\mathbf{M}_{\mathrm{top}}$) there, which increases $E_{\mathrm{z}}$. At the same time, the presence of many domain walls would add substantial domain wall energy. Together, these costs make the domain state energetically unfavorable compared to the uniform state. This behavior can be understood from the schematic in Figure~\ref{fig:4}f. While the presence of skyrmions has been previously reported via MFM in a similar trilayer SAF system,\cite{sim2025zero} the absence of skyrmions in our trilayer SAF system agrees both with the above energy considerations and our micromagnetic modeling.

\begin{figure*}[t]
\centering
\includegraphics[width=1.0\textwidth]{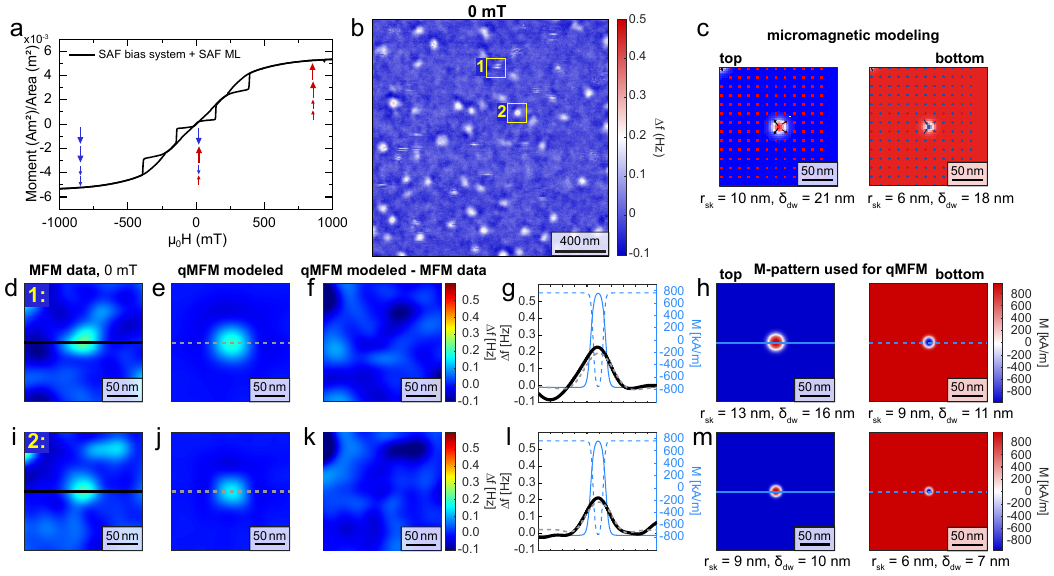}
\caption{a VSM hysteresis loop of the SAF bias system + SAF ML. b Background-subtracted MFM data of the SAF bias system + SAF ML at 0\,mT. c Micromagnetic simulation of the topmost and bottommost layers of the SAF ML when above the SAF bias system revealing an ultra-small SAFsk stabilized at zero field. Analysis of skyrmion 1 (d–h) and skyrmion 2 (i–m) in the SAF bias system + SAF ML system: d,i Background-subtracted MFM data of the skyrmions cut from the overview image at the positions highlighted by the yellow squares and labels 1 and 2. e,j Optimized qMFM simulations of the SAF skyrmion. f,k Residual map from subtracting the experimental from the modeled data. g,l Line profiles of the experimental (solid black lines) and modeled (dashed gray lines) MFM skyrmion frequency-shift signals and the skyrmion magnetization profiles for the top and bottom SAF layers (solid and dashed blue lines, respectively). h,m Optimized magnetization pattern used for the qMFM analysis of the skyrmion in the top and bottom trilayers of the SAF ML.}
\label{fig:5}
\end{figure*}

\subsection{SAF Multilayer on SAF bias system}
The partially uncompensated SAF bias system (Figure~\ref{fig:1}a) was combined with the fully compensated SAF ML to achieve an overall magnetically compensated system. The resulting hysteresis loop in Figure~\ref{fig:5}a closely resembles that of the SAF ML alone (Figure~\ref{fig:4}a), but additional small switching events originating from the SAF bias system are visible. Importantly, the magnetic moment at zero field again vanishes, indicating that the combined SAF bias/SAF ML system remains fully compensated. Although the SAF bias system by itself is not fully compensated, interfacial effects at the boundary between the SAF bias system and the SAF ML likely suppress its net moment in the combined structure (see Supporting Information Section 4).
Figure~\ref{fig:5}b shows the background-subtracted zero-field MFM data of the combined structure. Background subtraction is essential here, as the raw frequency-shift signal contains contributions arising from sample roughness and local variations of the magnetic moment, which would otherwise mask the weak skyrmion contrast (see Supporting Information Figure S3). In the processed image, numerous circular features become visible. These exhibit comparatively weak contrast, typically in the range of $0.07$–$0.25\,\mathrm{Hz}$, with most lying below $0.15\,\mathrm{Hz}$. Such reduced contrast, well below half of that observed for FMsk, is consistent with the expected response of SAF-coupled skyrmions. However, confirming their SAF nature requires quantitative analysis through micromagnetic modeling and qMFM simulations.

Figure~\ref{fig:5}c shows the micromagnetic modeling results, specifically the topmost and bottommost Co layers obtained from a micromagnetic simulation of the two antiferromagnetically coupled [Ru/Co/Pt]$_{\times 3}$ trilayers on top of the bias SAF (the remaining Co layers of the whole ML exhibit very similar magnetization textures). In contrast to the SAF ML alone, the presence of the SAF bias system stabilizes a skyrmion with a core that is magnetized north in the top three layers and south in the bottom three layers. The skyrmion in the top trilayer has a radius of $r_{\text{sk}} = 10~\text{nm}$ and a domain wall width of $\delta_{\text{dw}} = 21~\text{nm}$, whereas the skyrmion in the bottom trilayer is smaller, with $r_{\text{sk}} = 6~\text{nm}$ and $\delta_{\text{dw}} = 18~\text{nm}$. The smaller diameter of the bottom trilayer skyrmion arises from the RKKY exchange field from the bias layer, which directly acts on the bottom trilayer, while the skyrmion spin texture in the top trilayer can slightly relax. This size difference reduces the mutual compensation of the stray fields generated by the two oppositely magnetized skyrmion cores: the smaller bottom layer skyrmion produces a more rapidly decaying stray field, which, combined with its larger distance from the surface, leads to incomplete cancellation and therefore a stronger net MFM contrast.

To quantitatively model the measured MFM frequency-shift contrast of the SAFsk (Figure \ref{fig:5}d to m), we again begin with parametrized circularly symmetric magnetization patterns. In this case, two such patterns are required: one with a north-oriented core in the top [Ru/Co/Pt]$_{\times 3}$ trilayer, and one with a south-oriented core in the bottom trilayer (Figure~\ref{fig:5}h and m, left and right panels, respectively), reflecting the AFM coupling between the two trilayers. In principle, the skyrmion radii and domain wall widths of the two trilayers could be chosen independently, leading to a total of four free parameters. Guided by the results of our micromagnetic modeling (Figure~\ref{fig:5}c), which consistently predict a smaller skyrmion in the bottom trilayer, we constrained the ratio of the top to bottom skyrmion radii and domain wall widths to a fixed value of 1.5. This reduces the problem to two independent parameters, while retaining the essential asymmetry between the two antiferromagnetically coupled skyrmions.

The resulting optimized parameters are
$r_{\rm sk}^{\rm bottom} = 9$\,nm, $\delta_{\rm dw}^{\rm bottom} = 11$\,nm and
$r_{\rm sk}^{\rm top} = 13$\,nm, $\delta_{\rm dw}^{\rm top} = 16$\,nm
for the measured frequency-shift pattern shown in Figure~\ref{fig:5}d.
Similarly, for the skyrmion in Figure~\ref{fig:5}i, the optimized values are
$r_{\rm sk}^{\rm bottom} = 6$\,nm, $\delta_{\rm dw}^{\rm bottom} = 7$\,nm and
$r_{\rm sk}^{\rm top} = 9$\,nm, $\delta_{\rm dw}^{\rm top} = 10$\,nm.
The simulated and corresponding difference images in Figure \ref{fig:5}~panels e, j and f, k, respectively, demonstrate the excellent agreement between experiment and modeling, confirming that the reconstructed magnetization textures faithfully reproduce the measured MFM signal. This agreement is further corroborated by the cross-sections in Figure~\ref{fig:5}g and l, where the calculated (dashed gray) and experimental (solid black) $\Delta f$ profiles closely overlap. The solid and dashed blue curves in these panels represent cross-sections through the magnetization patterns of the top and bottom [Ru/Co/Pt]$_{\times 3}$ trilayers, respectively.
As was also observed for the skyrmions in the FM ML, the skyrmion diameter estimated directly from the measured $\Delta f$ maps significantly exceeds the true diameter extracted from the magnetization profiles. This broadening is even more pronounced for SAFsk due to their intrinsically smaller radii and the enhanced cancellation of their stray fields. The true SAFsk size is the smallest reported in a SAF system, with a minimum diameter of $d_{\text{sk}} = 12~\text{nm}$, substantially smaller than skyrmions in comparable multilayer systems.\cite{legrand2020room,sim2025zero,pham2024fast,juge2022skyrmions} This reduced size can be understood from the compensated nature of the SAF ML, where the vanishing net saturation magnetization suppresses dipolar contributions, such that skyrmion stability is governed primarily by the DMI. The observed SAFsk diameters are close to those predicted for DMI-dominated skyrmions,\cite{buttner2018theory} indicating that the system operates predominantly in the DMI-stabilized regime, with only minor residual dipolar contributions. These SAFsk were stable and consistently reproducible at zero field, regardless of the field history. Furthermore, the skyrmion core polarity can be selected by saturating the system in a south or north field before performing the measurement at zero field. When saturated in a south field, north-core skymrions are present in the top FM trilayer (south-core in the bottom trilayer) at zero field, whereas when saturated in a north field, south-core skymrions are present in the top FM trilayer (north-core in the bottom trilayer) at zero field. Details and MFM data showing this is in Supporting Information Section 5. Current-driven skyrmion dynamics in the SAF ML have been preliminarily investigated using micromagnetic simulations, revealing motion with a vanishing SkHE. These results demonstrate that the present SAF ML design supports current-driven motion at zero field, though further material optimization will be required to enhance the achievable velocities

%-----------------------------------------------------------
%-----------------------------------------------------------
%-----------------------------------------------------------

\section{Conclusions}
In summary, we establish a bias-engineered SAF as an effective route for stabilizing and detecting zero-field skyrmions in both FM and compensated SAF multilayers. By combining high-sensitivity MFM with background-subtracted data processing and quantitative MFM analysis supported by micromagnetic modeling, we reveal nanoscale spin textures with nanometer precision and determine their true magnetic length scales.
This approach enables the direct observation of ultra-small SAFsk at zero field, reaching diameters below 20 nm and as small as 12 nm, representing the smallest SAF skyrmions reported to date. The SAF bias architecture not only stabilizes these extremely compact textures but also provides deterministic polarity control through a simple preparatory field cycle.
These results demonstrate that SAF-bias engineering offers a powerful and scalable pathway toward zero-field operation, polarity control, and substantial size reduction in chiral spin textures. Together with the quantitative MFM framework developed here, this work lays the foundation for high-density, energy-efficient skyrmion technologies based on compensated magnetic multilayers.

\section{Experimental Section}
\threesubsection{Material Growth and Characterization} 
Samples were prepared in an AJA Orion sputter chamber, which maintained a base pressure of $\sim 1 \times 10^{-9}$~mbar. The DC magnetron sputtering was performed on thermally oxidized Si(100) substrates, in an Ar atmosphere of 2~$\mu$bar. For all systems, 50~\AA\ Ta and 20~\AA\ Pt were initially deposited serving as seed layers and 20~\AA\ Pt was used as a final capping layer to protect against oxidation. Layer thicknesses were calibrated using a quartz crystal monitor. Co was sputtered at a power density of 2.96~W/cm$^{2}$, while Pt, Ru, and Ta were sputtered at 0.99~W/cm$^{2}$. The corresponding growth rates were 0.32~\AA/s for Co, 0.23~\AA/s for Pt, 0.17~\AA/s for Ru, and 0.135~\AA/s for Ta.
All hysteresis loops were measured via VSM using a Lakeshore 8604 VSM with a maximum magnetic field of 2.03 T, in which both in-plane and out-of-plane measurements can be performed.

\threesubsection{MFM Measurements and Quantitative MFM analysis}
MFM experiments were conducted using a custom-built high-vacuum system operated at approximately \(1 \times 10^{-6}\)~mbar, capable of generating an in-situ magnetic field of up to \( 380\)~mT. By operating the cantilever under vacuum conditions, it is possible to achieve a mechanical quality factor of up to $Q \approx 1.5$~million, which significantly improves measurement sensitivity compared to air-based MFM and allows the application of a thin magnetic coating on the tip to cause minimal perturbations to the sample’s micromagnetic configuration.\cite{feng2022magnetic}
For the experiments, we employed an uncoated SS-ISC silicon cantilever from Team Nanotech GmbH with a nominal tip radius below 5~nm. The cantilever has a length of 225~\(\mu\)m and a width of 35~\(\mu\)m. Its resonance frequency was experimentally measured to be 59.337~kHz. Using the measured resonance frequency together with the cantilever’s geometrical dimensions, the spring constant was determined to be 1.39 N/m.\cite{hutter1993calibration,butt1995calculation}
To make the cantilever tip responsive to magnetic interactions, a 6.5~nm Co layer was sputter-deposited at room temperature onto a 2~nm Ta seed layer. The cantilever was driven at its resonance frequency using a Zurich Instruments phase-locked loop system, maintaining a constant oscillation amplitude of 10~nm, while the resulting frequency-shifts due to the tip-sample force gradient were recorded. Measurements were performed in a non-contact mode with capacitive feedback control.\cite{zhao2018magnetic}

\threesubsection{Quantitative MFM Modeling}
Quantitative modeling of the MFM frequency-shift contrast was performed by computing the stray fields generated by candidate magnetization textures and converting them into simulated $\Delta f$ maps using the experimentally calibrated tip transfer function, together with the stiffness, resonance frequency, and oscillation amplitude of the cantilever.\cite{feng2022quantitative} For the maze domain state of the FM ML, the initial magnetization map was obtained by discriminating the background-subtracted MFM image (Figure~\ref{fig:2}b), producing a binary up/down distribution $M_z(x,y)=\pm M_{\rm s}$, where $M_{\rm s}$ was obtained from VSM. For skyrmions, the starting configuration consisted of a binary cylindrical domain of radius $r$. At this stage, both classes of magnetization patterns contain only $\pm M_{\rm s}$ regions and do not yet include finite-width domain walls; these are introduced during Fourier-space processing as described below.

All calculations were performed in two-dimensional Fourier space, transforming $(x,y,z)$ into $(k_x,k_y,z)$. This formulation greatly simplifies the evaluation of stray fields from multilayer samples and enables efficient convolution with the calibrated tip response, including the finite cantilever oscillation amplitude along an axis tilted by approximately $12^\circ$ relative to the surface normal. Domain walls were incorporated by applying the Fourier-space transfer function of a Bloch-type $180^\circ$ wall with out-of-plane profile $M_z(x)=M_{\rm s}\tanh(x/\delta_{\mathrm{dw}})$, leading to the wall transfer function $\mathcal{T}_{\mathrm{dw}}(k)=k\,(\delta_{\mathrm{dw}}/2)/\sinh(k\delta_{\mathrm{dw}}/2)$. In the Ru/Co/Pt ML used here, the interfacial DMI enforces N\'eel-type domain walls with the same $M_z$ profile and width but with a magnetization vector rotation perpendicular to the wall direction (clockwise for the negative DMI of this system). This magnetization rotation generates magnetic volume charges inside the domain wall, obtained from $-\nabla\!\cdot\!\mathbf{M}_{\mathrm{N\acute{e}el}}$.

These volume charges are combined with the surface charges $\sigma = {\bf M}\cdot \hat{\bf n}$ arising from the out-of-plane magnetization to form the effective charge density $\sigma_{\rm eff}({\bf k})=\sigma({\bf k})+\nabla_{\!\bf k}\cdot\mathbf{M}({\bf k})/k$, which serves as the source term for the stray field calculation. $\nabla_{\!\bf k} = (ik_x, ik_y, -k)$ is the nabla operator in Fourier space. Using $\sigma_{\rm eff}$, the $z$-component of the stray field above the multilayer surface and its derivative can be written analytically as
\begin{eqnarray}
H_z({\bf k},z) &=& \frac{1}{2}\,\sigma_{\rm eff}({\bf k}), 
e^{-kz}\, e^{-k t_{\rm CL}} \,[1 - e^{-k t_{\rm Co}}] 
\sum_{i=1}^{6} e^{-k t_i},\\
\frac{\partial H_z}{\partial z}({\bf k},z) &=& -k\,H_z({\bf k},z),
\end{eqnarray}
where $z$ is the tip–sample distance, $t_{\rm CL}=26$ {\AA} is the total thickness of the Ru/Pt capping layer, $t_{\rm Co}=0.6$ nm is the Co layer thickness, and $t_i$ denotes the depth of each magnetic layer relative to the surface.

The resulting $H_z({\bf k},z)$ or $\partial H_z/\partial z$ is multiplied by the calibrated tip transfer function\,\cite{feng2022quantitative} and by transfer functions accounting for the finite cantilever oscillation amplitude (approximately \ldots\,nm) along its tilted oscillation axis. After inverse Fourier transformation, this yields the expected $\Delta f(x,y)$ pattern for the candidate magnetization texture.

Finally, the free parameters of the model, primarily the skyrmion radius $r$ and the domain wall width $\delta_{\mathrm{dw}}$, were optimized by minimizing the rms difference between the simulated and measured $\Delta f$ maps. This approach yields the quantitatively reconstructed spin textures shown in Figure~\ref{fig:2}-\ref{fig:5}, providing sub-nanometer precision for $\delta_{\mathrm{dw}}$ and accurate determination of skyrmion radii in both the FM and SAF ML.

\threesubsection{Micromagnetic Simulations}
Micromagnetic simulations are performed by means of PETASPIN,\cite{giordano2012semi,xu2021imaging} a state-of-the-art CUDA-native in-house micromagnetic solver. The magnetic dynamics of the system is obtained integrating numerically the Landau-Lifshitz-Gilbert (LLG) equation by using the Adams-Bashforth scheme: 
\begin{equation}\label{eq:LLG_micro}
\frac{\partial\mathbf{m}}{\partial \tau} =
- \frac{1}{1+\alpha_\mathrm{G}^2} \, \left[ \mathbf{m} \times \mathbf{h}_{\mathrm{eff}} + \alpha_\mathrm{G} \mathbf{m}\times(\mathbf{m} \times \mathbf{h}_{\mathrm{eff}}) \right] ,
\end{equation}
where $\mathbf{m}=\mathbf{M}/M_\mathrm{s}$ is the normalized magnetization vector with $M_\mathrm{s}$ the saturation magnetization, $\tau = \gamma_0 M_\mathrm{s} t$ is the dimensionless time, with $\gamma_0$ the gyromagnetic ratio, and $\alpha_\mathrm{G}$ is the Gilbert damping constant. $\mathbf{h}_{\mathrm{eff}} = \mathbf{H}_{\mathrm{eff}}/(\mu_0 M_\mathrm{s})$ is the normalized effective field in units of $\mu_0 M_\mathrm{s}$, where $\mu_0$ is the vacuum permeability, and includes exchange, anisotropy, interfacial DMI, magnetostatic, and external fields.\cite{mandru2020coexistence} 

\begin{table}[h]
    \centering
    \caption{Micromagnetic parameters.\\ * Extracted from experimental data. ${\dagger}$ Estimated.}
    %\footnotesize * Extracted from experimental data. ${\dagger}$ Estimated.
    \label{table:mag_parameters_micro}
    \begin{tabular}{ccccc}
    \toprule  % Booktabs for a nicer top line
    System & $M_\mathrm{s}^{*}$  & $K_\mathrm{u}^{*}$  & $D^{\dagger}$  & $A^{\dagger}$ \\
    & (kA/m) & (MJ/m$^{3}$) & (mJ/m$^{2}$) & (pJ/m)\\
    \midrule
    FM ML & 920 & 0.68 & 1.4 & 8 \\
    SAF ML & 932 & 0.76 & 1.4 & 8\\SAF bias system & 1243 & 1.38 & 0.7 & 8\\
    \bottomrule
    \end{tabular}
    \vspace{2mm} 
    \parbox{0.9\linewidth}{
    \footnotesize
}
\end{table}

The effective field also includes the RKKY contribution, coupling the FM layers via indirect exchange across the metallic spacers. The $\mathbf{H}_{\mathrm{RKKY}}$ between two consecutive FM layers $i$ \text{and} $j$ is given by:

\begin{equation} \label{eq:RKKY_micro}
    \mathbf{H}_{\mathrm{RKKY},i(j)}= \frac{- J_\mathrm{RKKY} }{M_{s,i(j)} t_\mathrm{FM}} \mathbf{m}_{j(i)} ,
\end{equation}

where $t_\mathrm{FM}$ is the thickness of the discretization cell, that here corresponds to the FM layer thickness except for the SAF bias layer, and $J_\mathrm{RKKY}$ is the RKKY coupling strength between the two FM layers. 
The FM and SAF ML are simulated by six repetitions of a 6~\AA\ Co layer with a 30~\AA\ $\times$ 30~\AA\ $\times$ 6~\AA\ discretization cell. For the Co layers with FM coupling between them, the spacer was discretized along the z-direction with three cells due to the increased Ru thickness. For Co layers with AFM coupling, the spacer of Pt/Ru was discretized along the z-direction with one or two cells (when without or with Pt, respectively) due to the thinner Ru thickness. After a systematic study as a function of the RKKY strength, ${J_\mathrm{RKKY}}$ was fixed to 0.25 mJ/m$^{2}$ for all six repetitions of the FM ML, whereas it was fixed to -0.40 mJ/m$^{2}$ only for the AFM coupling between the two trilayers in the SAF ML. 
The SAF bias system is simulated with two repetitions of a 12~\AA\ Co layer separated by two non-magnetic layers. The AFM coupling both with the SAF bias system and between the SAF bias system and the ML on top was simulated with a RKKY strength, ${J_\mathrm{RKKY}}$ = -0.15 mJ/m$^{2}$.
We simulated a 300~nm~$\times$~300~nm square sample at zero temperature. The parameters used are shown in Table \ref{table:mag_parameters_micro}, $M_\mathrm{s}$ and $K_\mathrm{u}$ were calculated from the in-plane (not shown/Supporting Information Figure S2) and out-of-plane VSM measurements, and reasonable values for $D$ and $A$ were used.

% Acknowledgements
\medskip
\noindent
\textbf{Acknowledgments} \par %delete if not applicable))
The work performed at Empa was supported through the SNF Lead Agency project 200021E\_211828 ``Binary information represented by chiral spin textures".
The micromagnetic modeling work was supported by the project PRIN20222N9A73 ``SKYrmion-based magnetic tunnel junction to design a temperature SENSor-SkySens", funded by the Italian Ministry of University and Research (MUR). MR and RT acknowledge the support from the project PE0000021, ``Network 4 Energy Sustainable Transition - NEST", funded by the European Union - NextGenerationEU, under the National Recovery and Resilience Plan (NRRP), Mission 4 Component 2 Investment 1.3 - Call for Tender No. 1561 dated 11.10.2022 of the Italian MUR (CUP C93C22005230007). GF, MC, and RT are with the PETASPIN team and thank the PETASPIN Association (www.petaspin.com). 

\medskip
\noindent
\textbf{Conflicts of Interest} \par
The authors declare no conflict of interest.

\medskip
\noindent
\textbf{Data Availability Statement} \par
The data that support the findings of this study are available from the corresponding author upon reasonable request.

% References
\medskip

% Use the following code if you wish to generate your bibliography with BibTeX;
% replace the string "MSP-template" below with the name(s) of
% the BibTeX data base(s) you want to use.
% The resulting bibliography-output (the content of the .bbl file)
% must be pasted back into this file before submission.
% Please also include your BibTeX data base file(s) in your submission
% so that we can re-run BibTeX if necessary.
%

\bibliographystyle{MSP}
\bibliography{bib.bib}

\end{document}